\documentclass[a4paper, amsfonts, amssymb, amsmath, reprint, showkeys, nofootinbib, superscriptaddress, twoside]{revtex4-1}
\usepackage[english]{babel}
\usepackage[utf8]{inputenc}
\usepackage{CJK}
\usepackage[colorinlistoftodos, color=green!40, prependcaption]{todonotes}
\usepackage{amsthm}
\usepackage{mathtools}
\usepackage{physics}
\usepackage{xcolor}
\usepackage{graphicx}
\usepackage[left=23mm,right=13mm,top=35mm,columnsep=15pt]{geometry} 
\usepackage{adjustbox}
\usepackage{placeins}
\usepackage[T1]{fontenc}
\usepackage{lipsum}
\usepackage{csquotes}
\usepackage[pdftex, pdftitle={Article}, pdfauthor={Author}]{hyperref} 
\begin{document}
\begin{CJK}{UTF8}{gbsn}
\title{Elliptic anisotropy of open-charm hadrons from parton scatterings in p--Pb collisions at energies available at the CERN Large Hadron Collider}

\author{Siyu Tang}
    \affiliation{School of Mathematical and Physical Sciences, Wuhan Textile University, Wuhan 430200, China}
    \affiliation{Hubei Key Laboratory of Digital Textile Equipment, Wuhan Textile University, Wuhan 430200, China}
\author{Yuan Lu} 
    \affiliation{School of Electronic and Electrical Engineering, Wuhan Textile University, Wuhan 430200, China}
\author{Chao Zhang}
    \email[Correspondence email address: ]{chaoz@whut.edu.cn}
    \affiliation{School of Science, Wuhan University of Technology, Wuhan, 430200, China}
\author{RenZhuo Wan}
    \email[Correspondence email address: ]{wanrz@wtu.edu.cn}
    \affiliation{Hubei Key Laboratory of Digital Textile Equipment, Wuhan Textile University, Wuhan 430200, China}
    \affiliation{School of Electronic and Electrical Engineering, Wuhan Textile University, Wuhan 430200, China}
\date{\today} 

\begin{abstract}
The elliptic azimuthal anisotropy coefficient ($v_{2}$) of open-charm hadron at midrapidity ($|\eta<1|$) was studied in p--Pb collisions at $\sqrt{s_{\mathrm{NN}}}=$ 8.16 TeV using a multi-phase transport (AMPT) model included an additional charm quark–-antiquark pair production trigger. The model provides a simultaneously description of the measured $p_{\mathrm{T}}$ spectrum and $v_{2}$ of $D^{0}$ meson, as well as the $v_{2}$ of light flavor meson $K_{s}^{0}$. We found that the $D^{0}$ and $K_{s}^{0}$ $v_{2}$ are both significantly affected by different parton scatterings among charm and light quarks.  
In addition, the predictions for the $v_{2}$ of other charm hadrons including $D^{+}$, $D_{s}^{+}$ and $\Lambda_{c}^{+}$ in p--Pb collisions are given for the first time. The $v_{2}$ of open-charm hadron reasonably follows the number-of-constituent-quark (NCQ) scaling up to 2.5 GeV, strongly indicating the importance of partonic degrees of freedom for the collectivity of heavy flavors in high-multiplicity p--Pb collisions. These findings may hint the formation of deconfined matter in small collision systems, and provide referential value for future measurements of azimuthal anisotropy at the LHC energies. 
\end{abstract}

\maketitle
\section{Introduction}
The heavy-ion collisions at ultra-relativistic energies aim to investigate the properties of nuclear matter under extremely high temperatures and energy densities, known as the quark-gluon plasma (QGP)~\cite{Shuryak:1978ij,Shuryak:1980tp}. The QGP exhibits the behavior of a nearly perfect fluid, with a low shear viscosity to entropy density ratio~\cite{Qin:2010pf,Teaney:2010vd}, $\eta/s$. The study of the azimuthal anisotropy of final-state particles produced in heavy-ion collisions is a significant approach to constrain the transport properties of the QGP~\cite{Ollitrault:1992bk,Voloshin:2009fd}. This anisotropy is characterised in terms of the Fourier coefficients $v_n=<\mathrm{cos}[n(\varphi-\Psi_{n})]>$, where the $\varphi$ is the azimuthal angle of the final-state particle angle and $\Psi_{n}$ is the symmetry-plane angle in the collision for the $n$-th harmonic~\cite{Voloshin:1994mz,Poskanzer:1998yz}. The second order coefficient, $v_2$, referred to the elliptic flow, is connected to the almond-shaped overlap area formed by colliding nuclei and, as a result, constitutes the primary source of anisotropy in non-central collisions. 

Heavy quarks (charm and beauty) predominantly originate from initial hard-scattering processes characterised by timescales shorter than the QGP formation time, typically around 0.1$\sim$1 fm/$c$, therefore they experience the whole evolution of the QGP, and interact with the constituents of QGP medium~\cite{vanHees:2005wb,Brambilla:2010cs,Andronic:2015wma}. Such interactions accompanied by the energy loss lead to a strong modification of the open heavy-flavor hadron (i.e. mesons and baryons that carry one charm or bottom quark/anti-quark) yield in heavy-ion collisions with respect to pp collisions, which are widely observed in experiments from the Relativistic Heavy Ion Collider (RHIC) and the Large Hadron Collider (LHC)~\cite{Gyulassy:2004zy,STAR:2005gfr,PHENIX:2004vcz,Muller:2012zq}. Further understanding to the interactions of heavy quarks with QGP medium can be gained by analyzing the elliptic flow, $v_{2}$ of open heavy-flavour hadrons in heavy-ion collisions. Recent measurements at RHIC~\cite{STAR:2017kkh,STAR:2023eui} and the LHC~\cite{ALICE:2020iug,Grosa:2017zcz} show that the non-strange D-meson $v_2$ at low $p_{\mathrm{T}}$ is lower than that of pions and protons, following the the hypothesis of a mass hierarchy. It indicates that the charm quarks participate in the collective expansion of the medium, as well as undergo recombination with flowing light quarks. Additionally, studying the $v_{2}$ of D mesons with strange-quark content ($D_{s}$) is also very interesting as it allows further investigation into the effect of charm quark hadronization on D-meson flow~\cite{Grosa:2017zcz}.

In recent years, the flow-like phenomena of heavy flavors are also observed in small collision systems. The first measurement of elliptic azimuthal anisotropies for prompt $D^{0}$ mesons performed by the CMS collaboration, indicates that the $D^{0}$ $v_{2}$ has a sizable value and is lower than light-flavor hadron results~\cite{CMS:2018loe}. The ALICE collaboration has measured a significant $v_{2}$ of electrons and muons from heavy-flavor hadron decays in p--Pb collisions at mid and forward rapidities~\cite{ALICE:2018gyx,ALICE:2015lpx,ALICE:2022ruh}. However the origin of such collectivity of heavy flavors in small collision systems is still debated. In hydrodynamics-based models, a sizable $v_{2}$ is expected to result from significant interactions between charm quarks and the QGP medium, but it also generates the suppression of particle yield at high transverse momentum ($p_{\mathrm{T}}$) region, in contradistinction to the observed unity of $R_{\mathrm{pPb}}$ for charm hadrons~\cite{Du:2018wsj,Cao:2020wlm,Xu:2015iha}. The Color-Glass Condensate (CGC) calculations within the dilute-dense formalism considering the initial-state effect can reproduce the $v_{2}$ of non-prompt D mesons well in p--Pb collisions at mid rapidity~\cite{Zhang:2019dth,Zhang:2020ayy}, but it overestimates the data at forward rapidity~\cite{ALICE:2022ruh}. Recent studies about resolving the $R_{\mathrm{pA}}$ and $v_{2}$ puzzle of $D^{0}$ mesons with the transport models demonstrates the importance of both parton interactions and the Cronin effect in the high-multiplicity p--Pb collisions~\cite{Zhang:2022fum}. 

In addition, an approximate number-of-constituent-quark (NCQ) scaling of $v_{2}$ for light flavor hadrons was observed in high-multiplicity p--Pb collisions by the ALICE and CMS collaborations~\cite{CMS:2018loe,ALICE:2013snk}. It triggered the discussions about the existence of partonic degree of freedom in small systems. The viscous hydrodynamics combined with the linearized Boltzmann transport (LBT) model, including various hadronization mechanisms can well describe the identified particle $v_{2}$ at intermediate $p_{\mathrm{T}}$~\cite{Zhao:2020wcd}. Our previous studies~\cite{Tang:2023wcd} based on a multi-phase transport (AMPT) model demonstrated that the parton scatterings plus quark coalescence can also reproduce the NCQ scaling behavior for light flavor hadrons. However similar studies were still missing for heavy flavors in p--Pb collisions. Since the charm quarks are proved to be more hydrodynamic than light quarks in final-state azimuthal anisotropy~\cite{Li:2018leh}, probing the partonic collectivity for heavy quarks is crucial in searching possible formation of the QGP in the small systems at the LHC energies.
 
In this work, we incorporate an additional charm quark--antiquark($c\bar{c}$) pair trigger in the AMPT model to simultaneously describe the $v_{2}$ and $p_{\mathrm{T}}$ spectrum of $D^{0}$ mesons in high-multiplicity p--Pb collisions. The first study of the NCQ scaling of $v_{2}$ for open-charm hadrons, including $D^{0}$, $D^{+}$, $D_{s}^{+}$ and $\Lambda_{c}^{+}$, is performed over the $\it{p}_{\mathrm{T}}$ region from 0 to 10 GeV/$c$. We also investigate how the parton cascade mechanism implemented in AMPT affect elliptic anisotropy of charm hadrons in small collision systems. 

\section{The heavy-flavor triggered AMPT model}
In this analysis, we employed the string melting version of the AMPT model (v2.26t9b, available online)~\cite{Lin:2004en}, which has been demonstrated to successfully describe numerous observables at both RHIC and LHC energies. The AMPT is a hybrid framework that include four main processes: initial conditions, parton scatterings, hadronization and hadronic rescatterings. The initial conditions are handled by the heavy-ion jet interaction generator (HIJING) two-component model~\cite{Gyulassy:1994ew}, which explicates particle production in the context of both a soft and a hard component. The soft component is modeled by the formation of excited strings in non-perturbative processes, while the hard component involves the production of minijets from hard processes. In these hard processes, hard partons are produced with a momentum transfer larger than the cutoff momentum $p_0$, to regulates the total minijet production cross section, which can be expressed as: 
\begin{equation}
 \frac{\mathrm{d}\sigma^{cd}}{\mathrm{d}p_{\mathrm{T}}^{2}\mathrm{d}y_1\mathrm{d}y_2}=K\sum_{a,b}x_1f_a(x_1,p_{\mathrm{T}}^{2})x_2f_b(x_2,p_{\mathrm{T}}^{2})\frac{d\sigma^{ab \rightarrow Q\bar{Q}}}{d\hat{t}}
\label{minijet_cross_section}
\end{equation}
where $y_1$ and $y_2$ represent the rapidities of produced partons, $f_a$ and $f_b$ are the parton distribution function of parton type $a$ and $b$ in a nucleon, and $\sigma^{ab \rightarrow cd}$ is the cross section for parton $a$ and $b$ to produce the minijets $c$ and $d$. Then the total minijet cross section can be written as
\begin{equation}
\sigma_{\mathrm{jet}}=\sum_{c,d}\frac{1}{1+\delta_{cd}}\int^{s/4}_{p_{0}^2}dp_{\mathrm{T}}^{2}dy_1dy_2\frac{d\sigma^{cd}}{dp_{T}^{2}dy_1dy_2}
\label{Total minijet_cross_section}
\end{equation}
The differential cross section of heavy-quark pair in HIJING is evaluated by the perturbative quantum chromodynamics (pQCD) at leading order, which has the same form as Eq.~\ref{minijet_cross_section}: 
\begin{equation}
 \frac{\mathrm{d}\sigma^{Q\bar{Q}}}{\mathrm{d}p_{\mathrm{T}}^{2}\mathrm{d}y_1\mathrm{d}y_2}=K\sum_{a,b}x_1f_a(x_1,p_{\mathrm{T}}^{2})x_2f_b(x_2,p_{\mathrm{T}}^{2})\frac{d\sigma^{ab \rightarrow Q\bar{Q}}}{d\hat{t}}
\label{heavy_quark_cross_section}
\end{equation}
where the same minimum transverse momentum cut $p_0$ is used in calculating $\sigma^{Q\bar{Q}}$ as done in Eq.\ref{Total minijet_cross_section} for $\sigma_{\mathrm{jet}}$. However, as described in~\cite{Zheng:2019alz}, the heavy quark has large mass, which can naturally regulates the heavy quark total cross section. Therefore, the $p_0$ cutoff may result in an additional suppression of heavy quark production.For this issue, we employed an approach involving the introduction of $c\bar{c}$ trigger to significantly enhance the production rate of heavy quarks. This trigger algorithm was initially developed to produce dijet event in HIJING model~\cite{Gyulassy:1994ew}, then was extended to trigger $c\bar{c}$ pair productions ($q+\bar{q}\rightarrow Q+\bar{Q}, g+g\rightarrow Q+\bar{Q}$). Such approach was widely used in the study of heavy-flavor hadron in heavy-ion collisions using the AMPT model~\cite{Wang:2019vhg,Wang:2021xpv,Wang:2022fwq}, and it has been demonstrated to be analogous to a recently proposed extended AMPT version~\cite{Wang:2022fwq,Zheng:2019alz}, where the $p_0$ cutoff was removed for heavy quark production. 

In the string melting mechanism, the produced light and heavy quarks are converted into primordial hadrons based on the Lund fragmentation~\cite{Lin:2004en}. Two key parameters $a$ and $b$ are used to determine the Lund string fragmentation function as $f(z)\propto z^{-1}(1-z)^a e^{-bm_{\perp}^{2}/z}$, where $z$ is the light-cone momentum fraction of the produced hadron of transverse mass $m_{\perp}$ with respect to the fragmenting string. Then these primordial hadrons are converted into partons according to their flavor and spin structures, thus form a dense partonic matter. The evolution of the partonic matter was simulated using Zhang's Parton Cascade (ZPC) model~\cite{Zhang:1997ej}. This model incorporates two-body elastic scatterings with a cross-section defined by the simplified equation below:
\begin{equation}
\sigma_{gg} \simeq \frac{9\pi\alpha_{s}^{2}}{2\mu^{2}},
\label{zpc cross_section}
\end{equation}
where the $\alpha_{s}$ is the strong coupling constant, and $\mu$ is the Debye screening mass. After the partons stop scattering, the nearest two (or three) quarks are combined into mesons (or baryons) with a quark coalescence model. The subsequent hadronic rescattering processes are described by an extended relativistic transport (ART) model~\cite{Li:1995pra} including both elastic and inelastic scatterings for baryon-baryon, baryon-meson, and meson-meson interactions.

In this study, the string melting AMPT model with and without $c\bar{c}$ trigger are used. In both results, 12 million events are simulated for inclusive p--Pb collisions at $\sqrt{s_{\mathrm{NN}}} = 8.16$ TeV. We set the $\alpha_{s}$ to 0.33, and adjusted the parton cross section $\sigma$ ($\sigma = 0,0.2,0.5$mb) by varying the parton screening mass $\mu$. The Lund string fragmentation parameters $a$ and $b$ are set to 0.3 and 0.15 $\mathrm{GeV}^{-2}$, respectively.


\section{Results and Discussions}
\begin{figure}[!hbt]
\begin{center}
\includegraphics[width=1.0\columnwidth]{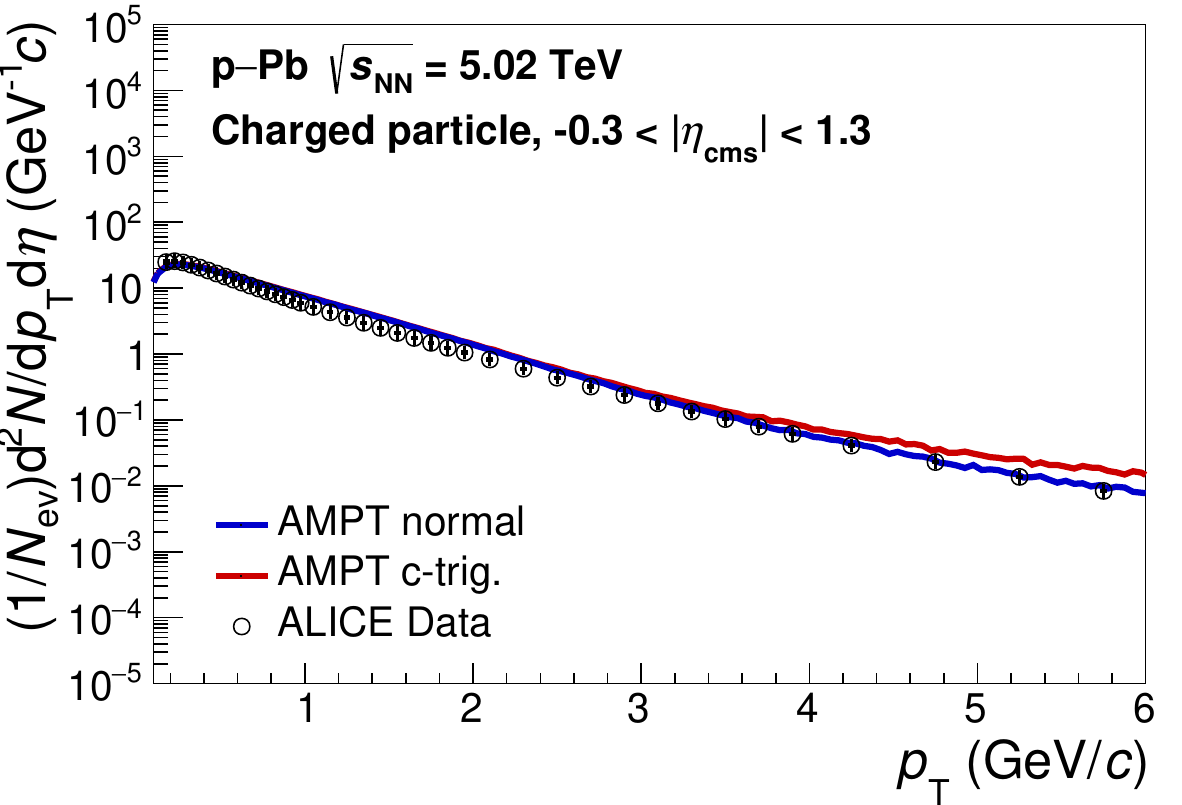}
\includegraphics[width=1.0\columnwidth]{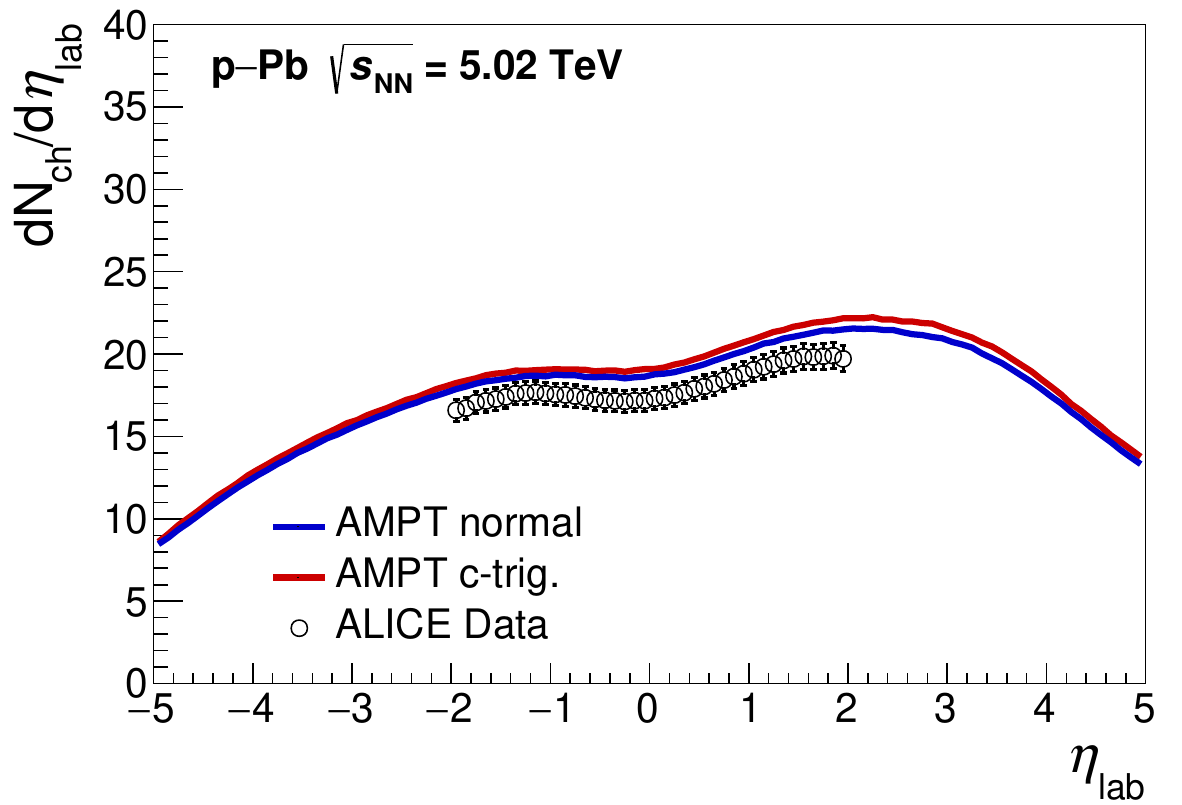}
\caption{(Color online) The $p_{\mathrm{T}}$ spectrum (top) and pseudorapidity density (bottom) of charged particles obtained from the AMPT model, are compared to the ALICE measurement~\cite{ALICE:2018vuu,ALICE:2012xs}.}
\label{Fig: charged spec}
\end{center}
\end{figure}

Before studying the elliptic anisotropy of the open-charm hadrons, we test the effect from the charm quark–-antiquark pair production trigger on the multiplicity distribution and $p_{\mathrm{T}}$ spectrum of the final-state charged particle. Figure~\ref{Fig: charged spec} shows the $p_{\mathrm{T}}$ spectrum and the pseudorapidity distribution ($dN_{\mathrm{ch}}/d\eta$) of charged particles in p--Pb collisions at $\sqrt{s_{\mathrm{NN}}}$ = 5.02 TeV obtained from the AMPT model with and without the additional $c\bar{c}$ trigger, and the comparisons with the ALICE measurement~\cite{ALICE:2018vuu,ALICE:2012xs}. Following the previous studies~\cite{Zhang:2022fum}, the parton cascade cross section in the AMPT is set to 0.5 mb. One can see that the results with $c\bar{c}$ trigger (labeled as "AMPT c-trig.") are slightly higher than that without $c\bar{c}$ trigger (labeled as "AMPT normal"), and both two sets of calculations provide reasonable descriptions of the data.   

\begin{figure*}[!hbt]
\begin{center}
\includegraphics[width=1.0\columnwidth]{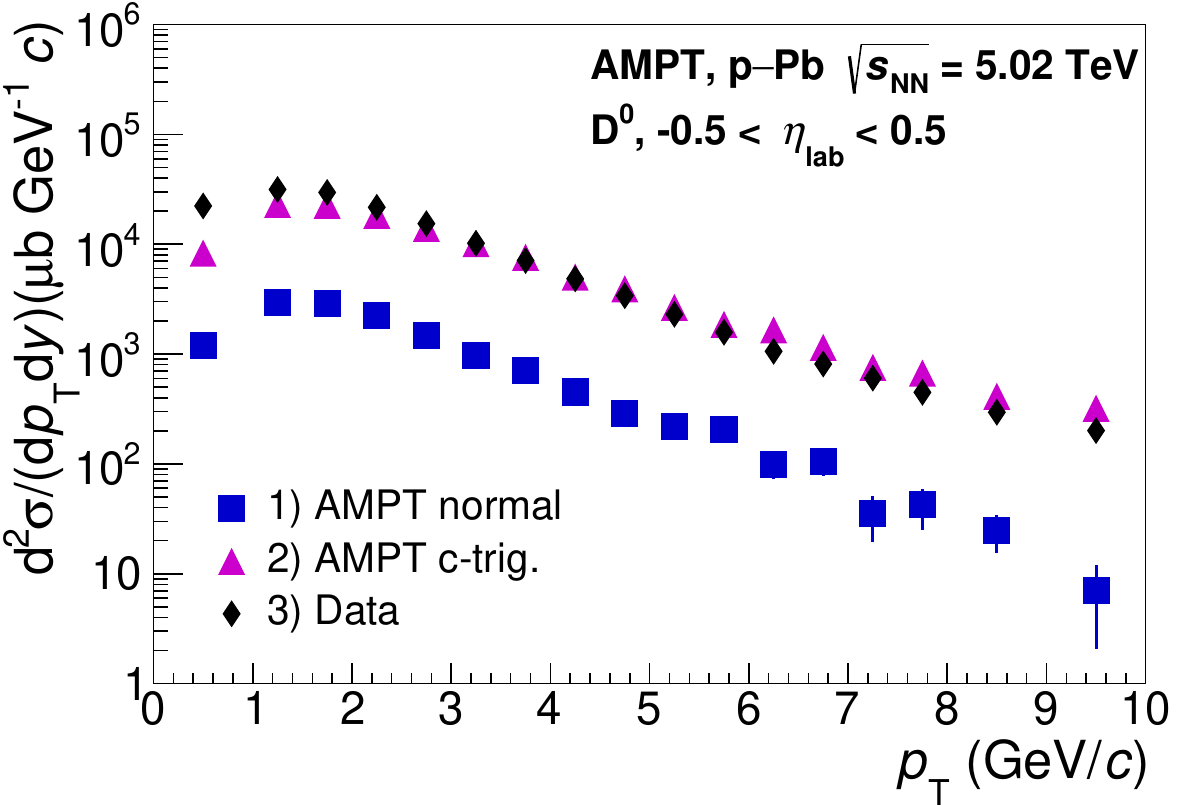}
\includegraphics[width=1.0\columnwidth]{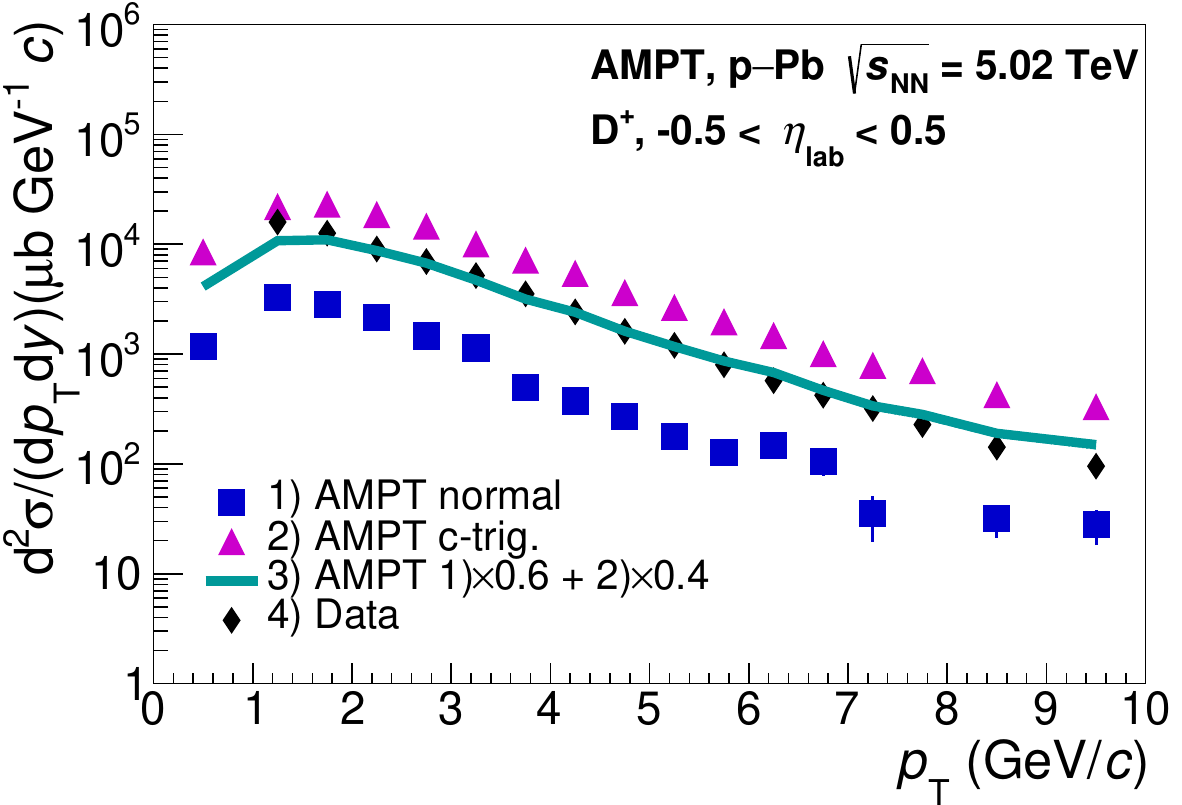}
\includegraphics[width=1.0\columnwidth]{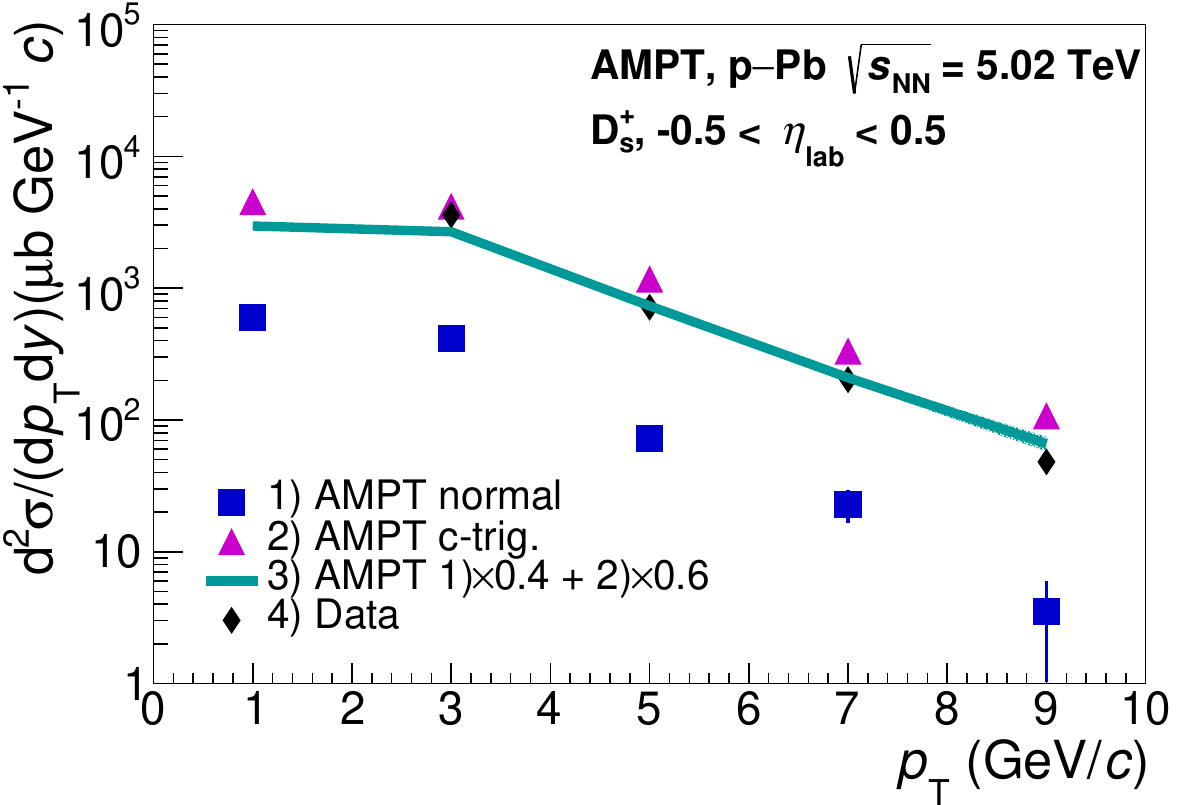}
\includegraphics[width=1.0\columnwidth]{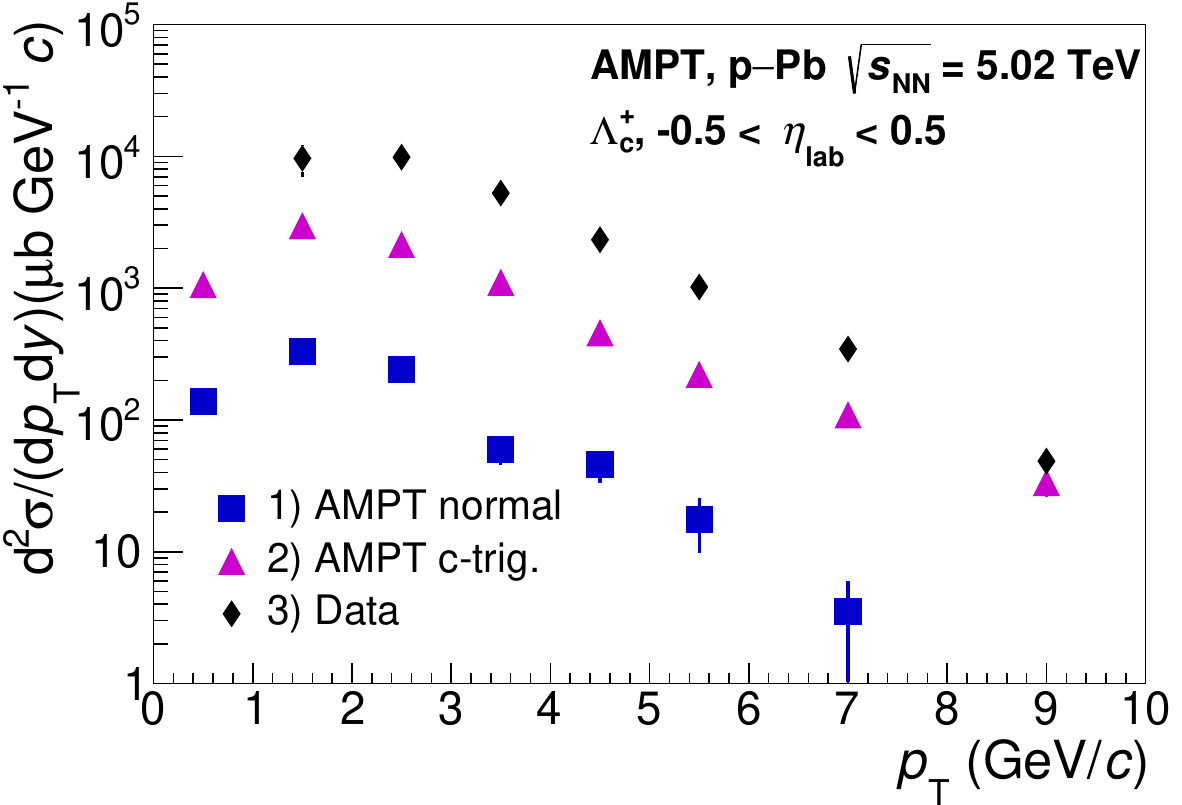}
\caption{(Color online) The cross section of $D^{0}$, $D^{+}$, $D_{s}^{+}$ and $\Lambda_{c}^{+}$ as a function of $p_{\mathrm{T}}$ in p--Pb collisions at 5.02 TeV obtained from AMPT model calculations for different configurations, are compared to the ALICE measurement~\cite{ALICE:2019fhe}.}
\label{Fig: D spec}
\end{center}
\end{figure*}

Figure~\ref{Fig: D spec} shows the $p_{\mathrm{T}}$ spectrum of open charm mesons ($D^{0}$, $D^{+}$, $D_{s}$) and baryons ($\Lambda_{c}^{+}$) from the AMPT model with different configurations in p--Pb collisions at $\sqrt{s_{\mathrm{NN}}}$ = 5.02 TeV in comparison with the ALICE measurement~\cite{ALICE:2019fhe,ALICE:2020wla}. The blue full squares represent the calculations from the "AMPT normal", which underestimate all measured charm hadron yields. The results from the "AMPT c-trig.", represented by violet full triangles, significantly enhanced the $p_{\mathrm{T}}$ spectrum compared to the "AMPT normal", indicating the higher generation rate for the open-charm hadron. Similar phenomena were also observed in previous studies in heavy-ion collisions~\cite{Wang:2021xpv}. From the comparisons with the data shown in Fig.~\ref{Fig: D spec}, one can see that the calculations from "AMPT c-trig." reasonably describe the measured $D^{0}$ yield, but overestimate the $D^{+}$ and $D_{s}^{+}$ spectrum. Since the $p_{\mathrm{T}}$ slops of these two AMPT calculations are same, we build two sets of new event samples to describe the $D^{+}$ and $D_{s}$ spectrum with the fractions determined by the data. The green lines shown in Fig~\ref{Fig: D spec}
represent the results from such new event samples, with 60\% of the "AMPT normal" plus 40\% of the "AMPT c-trig." for $D^{+}$ and 40\% of the "AMPT normal" plus 60\% of the "AMPT c-trig." for $D_{s}^{+}$. In addition, the data of $\Lambda_{c}^{+}$ $p_{\mathrm{T}}$ spectrum is still underestimated by the model calculations even the $c\bar{c}$ trigger is implemented. It may be ascribed to the coalescence model implemented in current AMPT version, where the baryons are formed only after the formation of meson by simply combining three nearest partons regardless of the relative momentum among the coalescing partons~\cite{He:2017tla}. In the following, the two new mixed samples are used to investigate the $v_{2}$ of $D^{+}$ and $D_{s}^{+}$, and the event sample from "AMPT c-trig." is used for $D^{0}$ and $\Lambda_{c}^{+}$. 

\begin{figure*}[!hbt]
\begin{center}
\includegraphics[width=1.9\columnwidth]{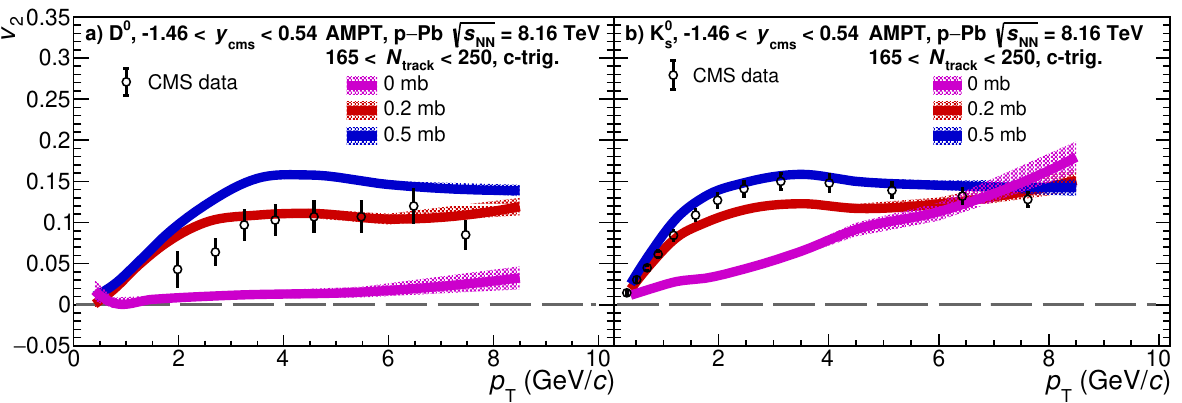}
\caption{(Color online) The $v_{2}$ of $K_s^{0}$ and $D^{0}$ as a function of $p_{\mathrm{T}}$ obtained from the AMPT model calculations with charm quark-anitquark trigger. The results with three set of parton cross section values are shown.}
\label{Fig: K0s_D0_v2}
\end{center}
\end{figure*}

\begin{figure*}[!hbt]
\begin{center}
\includegraphics[width=1.9\columnwidth]{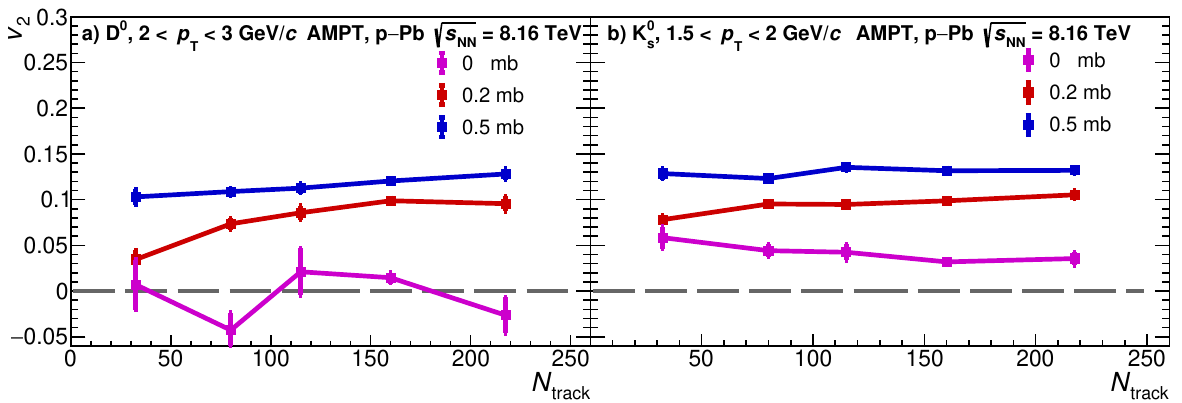}
\caption{(Color online) The $v_{2}$ of $K_s^{0}$ and $D^{0}$ as a function of $N_{\mathrm{track}}$ obtained from the AMPT model calculations with charm quark-anitquark trigger. The results with three set of parton cross section values are shown.}
\label{Fig: K0s_D0_v2_Cen}
\end{center}
\end{figure*}

In order to directly compare the calculations of $v_{2}$ with the data, we exactly follow the two-particle correlation method employed by the CMS experiments~\cite{CMS:2018loe}. The identified particles within the rapidity range -1.46 $<y_{\mathrm{cm}}<$ 0.54 are regarded as the trigger particles (denoted as "trig"), then they are correlated with the reference charged particles with $0.3 <p_{\mathrm{T}}<$ 3 GeV/$c$ in -2.4 $<\eta<$ 2.4 (denoted as "ref"). The azimuthal correlation distribution of these two emission particles can be expanded in the Fourier series as follows:
\begin{equation}
\frac{1}{N_{\mathrm{trig}}}\frac{\mathrm{d}N^{\mathrm{pair}}}{d\Delta \varphi} = \frac{N_{\mathrm{assoc}}}{2\pi}(1 + 2 \sum_{n=1}^{3}V_{n\Delta}(\mathrm{trig},\mathrm{ref})\mathrm{cos}(n\Delta\varphi)),
\label{eq: Fourier}
\end{equation}
where $\Delta\eta$ and $\Delta\varphi$ are the differences in $\eta$ and $\varphi$ of each particle pair, $V_{n\Delta}$ are the Fourier coefficients and $N_{\mathrm{assoc}}$ represents the total number of pairs per trigger particle. In order to suppress the nonflow contribution from the jet correlations, the $|\Delta\eta| > 1$ is applied in constructing such two-particle correlation. Assuming factorization of the Fourier coefficients, the $v_{2}$ of the trigger particles can be obtained by    
\begin{equation}
v_{2}(\mathrm{trig}) = \frac{V_{2\Delta}(\mathrm{trig},\mathrm{ref})}{\sqrt{V_{2\Delta}(\mathrm{ref},\mathrm{ref})}},
\label{eq: factorization}
\end{equation}
Figure~\ref{Fig: K0s_D0_v2}(a) and Fig~\ref{Fig: K0s_D0_v2}(b) show the $p_{\mathrm{T}}$-differential $v_{2}$ of $D^{0}$ and $K_{s}^{0}$ in p--Pb collisions at 8.16 TeV obtained from the AMPT calculations with the charm-quark trigger, and the comparisons with the CMS data~\cite{CMS:2018loe}. The high multiplicity events within 165 $<N_{\mathrm{track}}<$ 250 are selected\footnote{Note that it is a slightly looser selection compared to the cut 185 $<N_{\mathrm{track}}<$ 250 used in data~\cite{CMS:2018loe}, however its effect on the $v_{2}$ is negligible, as discussed in Fig.~\ref{Fig: K0s_D0_v2_Cen}}, where $N_{\mathrm{track}}$ is the number of charged particles with $p_{\mathrm{T}}>$ 0.4 GeV/$c$ within $|\eta|<$ 2.4. Three sets of parton cross section values are used in the model calculations. One can see that the $v_{2}$ of $K_{s}^{0}$ and $D^{0}$ has similar $p_{\mathrm{T}}$ trend and magnitude with the same and nonzero cross section value ($\sigma$= 0.2, 0.5 mb). And both $v_{2}$ increase with the increasing of cross section values. The $D^{0}$ $v_{2}$ obtained with a parton cross section of 0.2 mb provide a good description of data for $p_{\mathrm{T}} > 3$ GeV/$c$ while the results from 0.5 mb is systematically higher than the data. For $K_{s}^{0}$, the $v_{2}$ from 0.5 mb is closer to data compared to other $\sigma=0.2$ settings. Apparently the calculation with one fixed $\sigma$ can not provide a simultaneously description of the $v_{2}$ of $D^{0}$ and $K_{s}^{0}$. It suggests that the scattering probability among light quarks is higher than that between charm quark and light quarks, and they may need to be determined from data separately, which was also demonstrated in our previous work~\cite{Zhang:2022fum}. In addition, we calculate the $v_{2}$ of $D^{0}$ and $K_{s}^{0}$ when the parton scattering process is turn off ($\sigma=0$ mb), as the dash lines shown in Fig.~\ref{Fig: K0s_D0_v2}. A very small and finite value is obtained for the $D^{0}$ $v_{2}$, while the $K_{s}^{0}$ $v_{2}$ is significant and increase with the increasing $p_{\mathrm{T}}$. Similar phenomenon was observed in previous $v_{2}$ analysis in quark level~\cite{Zhang:2022fum}. It indicates that the charm hadron $v_{2}$ is mostly generated from the parton scatterings, while for the light flavor hadron, the contribution from the initial state correlation before parton scattering process (or nonflow) is not negligible especially at high $p_{\mathrm{T}}$ region. 

\begin{figure*}[!hbt]
\begin{center}
\includegraphics[width=1.9\columnwidth]{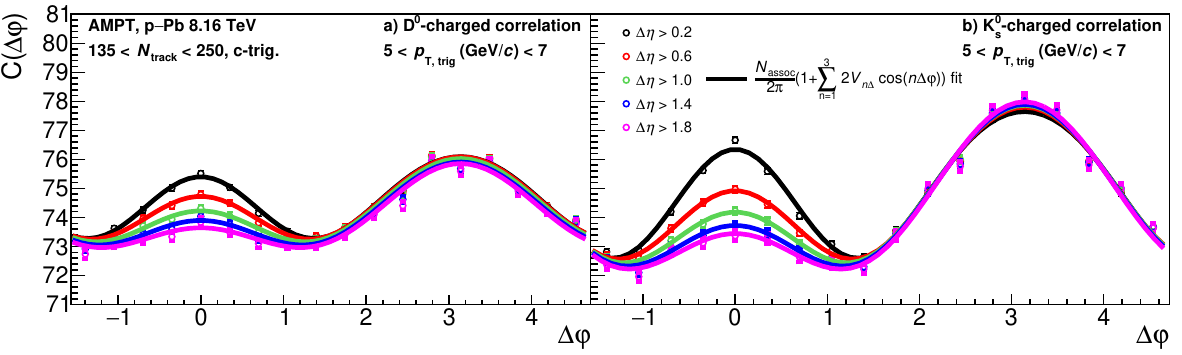}
\caption{(Color online) The distributions for $D^{0}$-charged (left) and $K_{s}^{0}$-charged (right) correlation with different $\Delta\eta$ cuts. The trigger particles are selected in 5 $<\it{p}_{\mathrm{T}}<$ 7 GeV/$c$.}
\label{Fig: Corr_cuts}
\end{center}
\end{figure*}

\begin{figure*}[!hbt]
\begin{center}
\includegraphics[width=0.95\columnwidth]{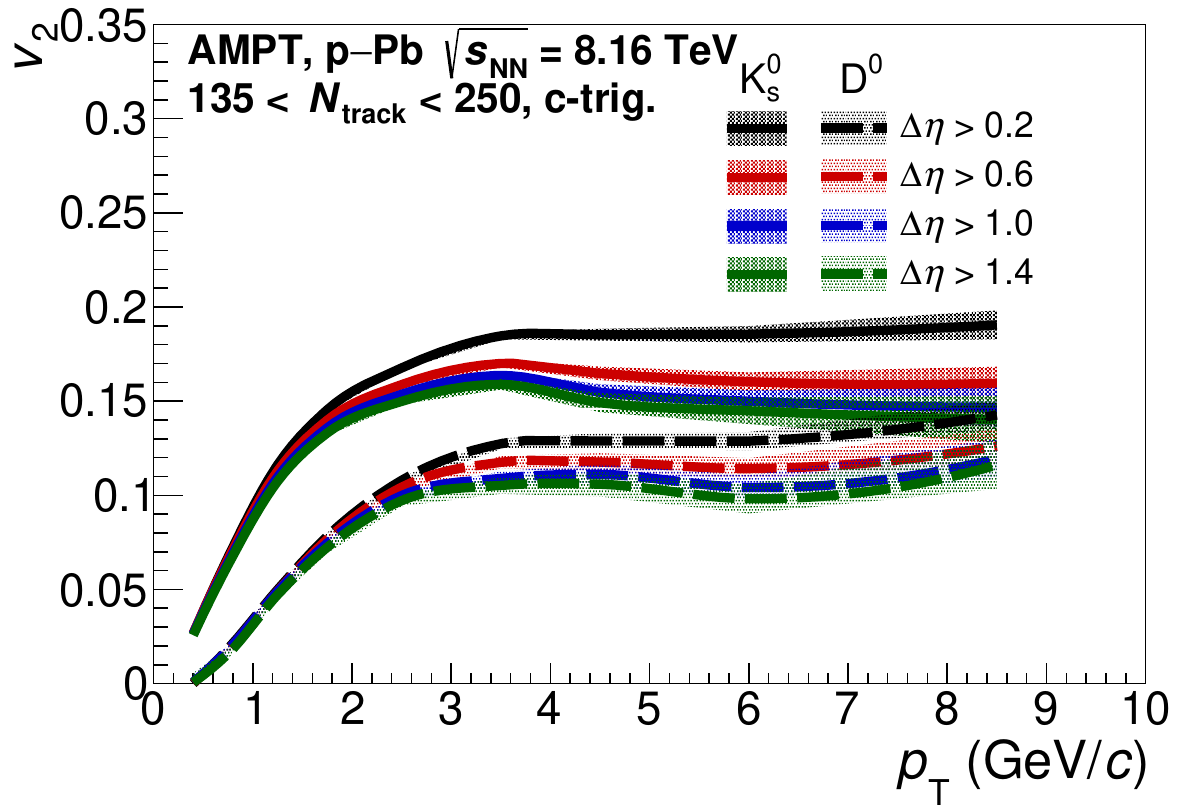}
\includegraphics[width=0.95\columnwidth]{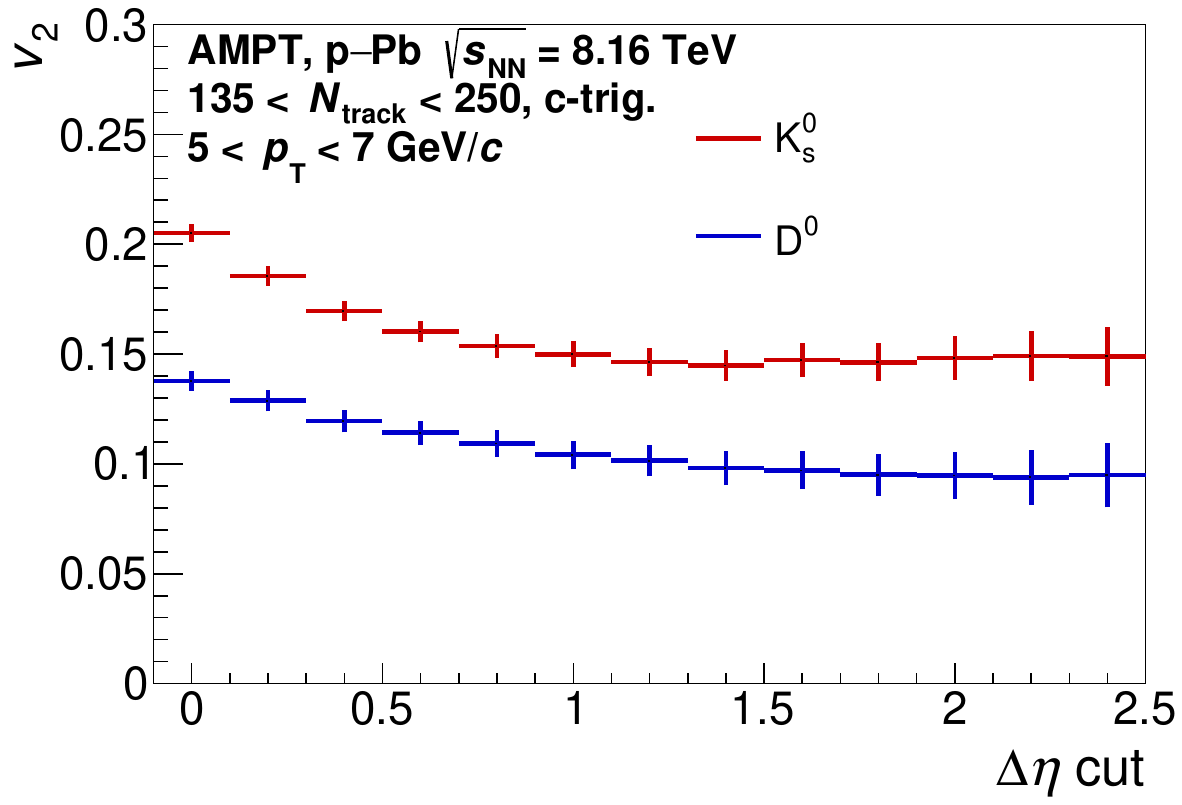}
\caption{(Color online) Left: the $v_{2}$ of $D^{0}$ and $K_{s}^{0}$ as a function of $\it{p}_{\mathrm{T}}$ with various $\Delta\eta$ cuts. Right: the $v_{2}$ of $D^{0}$ and $K_{s}^{0}$ as a function of $\Delta\eta$ cut.}
\label{Fig: pt_v2_deta}
\end{center}
\end{figure*}

Figure~\ref{Fig: K0s_D0_v2_Cen}(a) and Fig.~\ref{Fig: K0s_D0_v2_Cen}(b) present the $p_{\mathrm{T}}$-integrated $v_{2}$ of $K_{s}^{0}$ and $D^{0}$ as a function of $N_{\mathrm{track}}$. When the implemented cross section value is nonzero ($\sigma$=0.2, 0.5 mb), the $v_{2}$ values slightly increase from low-multiplicity to high-multiplicity events and same conclusions for $K^{\pm}$, $\pi^{\pm}$ and proton were obtained in our previous studies~\cite{Tang:2023wcd}. It indicates the larger elliptic anisotropy in more central collisions for both light- and heavy-flavor hadrons, reflecting the changing dynamic conditions and particle production mechanisms in different collision zones. On the other hand, when we exclude the parton scattering process ($\sigma$ = 0 mb), the $v_{2}$ of $K_{s}^{0}$ decreases with the increasing $N_{\mathrm{track}}$, while the $D^{0}$ $v_{2}$ fluctuates around 0. It is consistent with our findings about the $\it{p}_{\mathrm{T}}$-differential $v_{2}$ results in Fig.~\ref{Fig: K0s_D0_v2}, which hints a larger nonflow contribution to the $v_{2}$ of light quarks at low-multiplicity events compared to charm quarks.     

As described above, the nonflow contribution especially from the near-side jet correlation can be suppressed by introducing the pseudorapidity gap ($\Delta\eta > X$) in the two-particle correlation distribution (Eq.~\ref{eq: Fourier}). We vary the $\Delta\eta$ cut in a wide range (0$<X<$2.4) and test the stability of $v_{2}$ extraction for both $D^{0}$ and $K_{s}^{0}$ based on these cuts. To reflect the real data as closely as possible, the parton scattering cross section $\sigma$ is set to 0.2 mb for $D^{0}$, and 0.5 mb for $K_{s}^{0}$. Figure~\ref{Fig: Corr_cuts} shows the calculated $D^{0}$-charged (left) and $K_{s}^{0}$-charged (right) correlation distribution with various cuts ($X$= 0.2, 0.6, 1.0, 1.4, 1.8). One can clearly see that the near-side ($-0.5\pi<\Delta\varphi<0.5\pi$) correlation distribution is gradually reduced with the increase of the $\Delta\eta$ cut, indicating the subtraction of the near-side jet correlation. Figure~\ref{Fig: pt_v2_deta} (left) shows the $\it{p}_{\mathrm{T}}$-differential $v_{2}$ for $D^{0}$ and $K_{s}^{0}$ with these applied $\Delta\eta$ cuts. With the increase of $\Delta\eta$ cut, the $v_{2}$ for both $D^{0}$ and $K_{s}^{0}$ decrease, especially at high $\it{p}_{\mathrm{T}}$ region where the jet contribution is dominant. In addition, we also investigate the dependence of $v_{2}$ on $\Delta\eta$ cut, as shown in Fig.~\ref{Fig: pt_v2_deta} (right). One can see that the $v_{2}$ of $D^{0}$ and $K_{s}^{0}$ in 5 $<\it{p}_{\mathrm{T}}<$ 7 GeV/$c$ decreases with the increasing of the width of the introduced $\eta$ gap, but becomes almost flat for $\Delta\eta>1$. It indicates that the applied $\Delta\eta>1$ cut is reasonable to suppress the nonflow contribution, and reflects the maximum width of the near-side jet correlation in high-multiplicity p--Pb collisions.

\begin{figure}[hbt]
\begin{center}
\includegraphics[width=1.0\columnwidth]{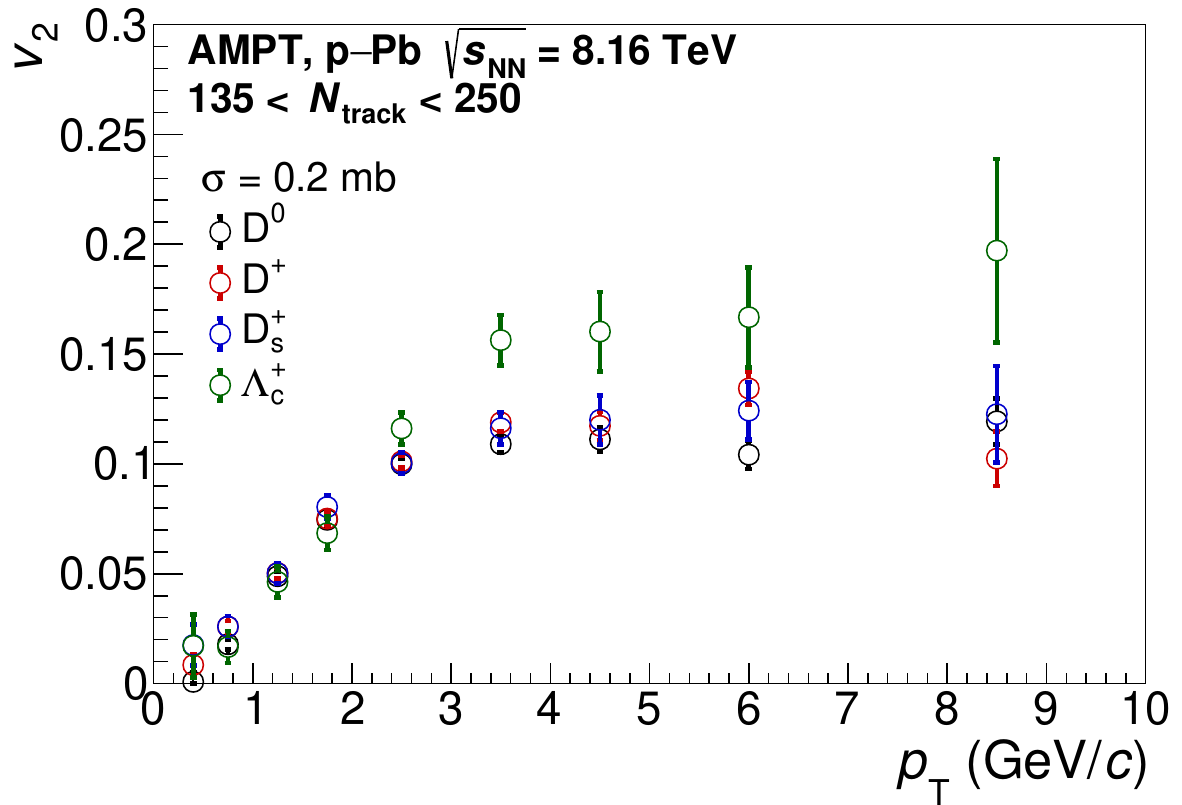}
\caption{(Color online) The $v_{2}$ of $D^{0}$, $D^{+}$, $D_{s}^{+}$ and $\Lambda_{c}^{+}$ as a function of $p_{\mathrm{T}}$ obtained from the AMPT model calculations with charm quark-anitquark trigger.}
\label{Fig: D_PID_v2}
\end{center}
\end{figure}

To further investigate the elliptic anisotropy of open-charm hadrons in small collision systems, we extend the calculations of $v_{2}$ to $D_{s}^{+}$, $D^{+}$ and $\Lambda_{c}^{+}$, as shown in the Fig.~\ref{Fig: D_PID_v2}. As discussed above, the AMPT model with charm quark-antiquark trigger is used for the $D^{0}$ and $\Lambda_{c}^{+}$ and mixed event samples introduced in Fig.~\ref{Fig: D spec} are used for $D^{+}$ and $D_{s}^{+}$. The parton scattering cross section value $\sigma$ is set to 0.2 mb, which is same as the $D^{0}$. One can see that the $v_{2}$ for all charm-hadron species is nonzero. The $v_{2}$ of $D^{0}$, $D^{+}$ and $D_{s}^{+}$ are consistent within uncertainties, which is compatible to the findings in heavy-ion collisions~\cite{ALICE:2020iug}. The $v_{2}$ of $\Lambda_{c}^{+}$ is larger than that of charm meson for $p_{\mathrm{T}} >$ 2 GeV/$c$, indicating that meson-baryon grouping behaviors also present in heavy flavor sector. Future measurements about the elliptic flow of charm baryons can provide more constrain on our calculations.   

\begin{figure}[hbt]
\begin{center}
\includegraphics[width=1.\columnwidth]{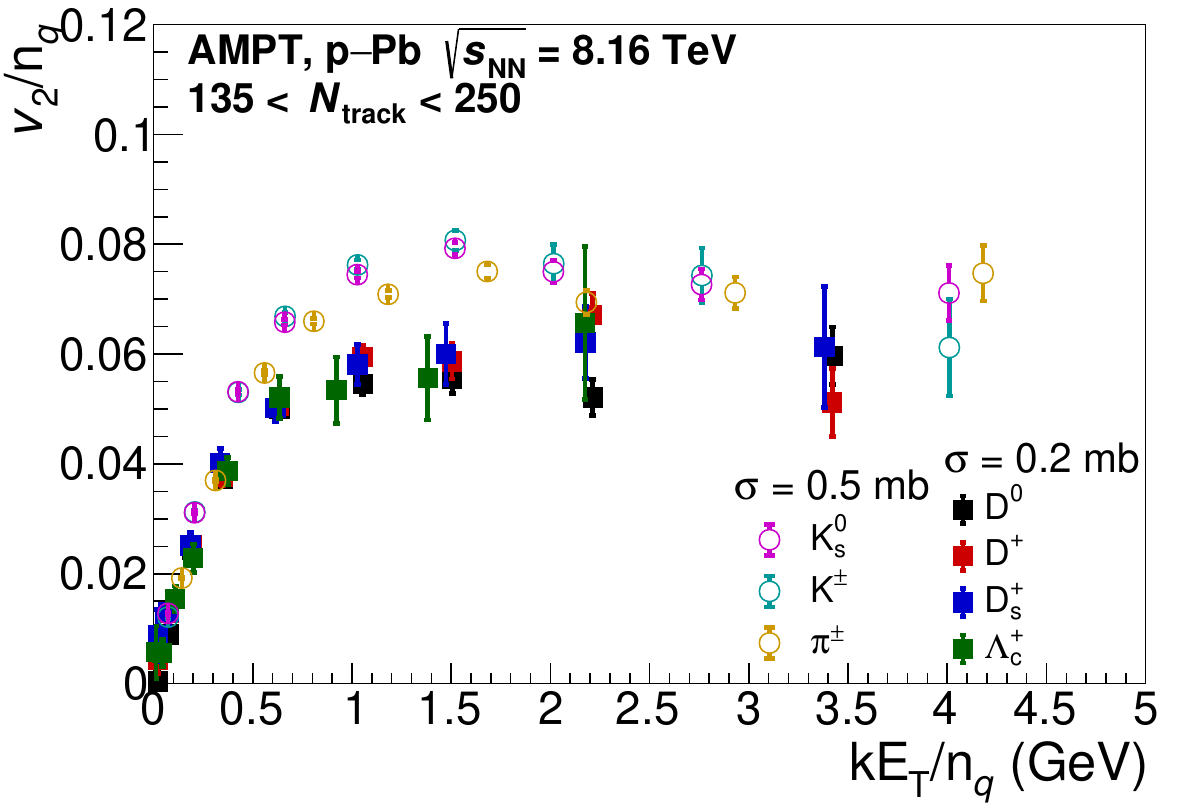}
\caption{(Color online) The $kE_{\mathrm{T}}$-differential $v_{2}$ of $D^{0}$, $D^{+}$, $D_{s}^{+}$ and $\Lambda_{c}^{+}$ scaled by the the number of constituent quark. The comparison to the results of $K_{s}^{0}$, $K^{\pm}$ and $\pi^{\pm}$ is also shown.}
\label{Fig: D_NCQ_v2}
\end{center}
\end{figure}

As discussed in Ref.~\cite{PHENIX:2007tef}, the observed meson-baryon particle type grouping for flow measurements in heavy-ion collisions indicates the collective behavior at the partonic level, which can be further studied by means of the NCQ scaling technique~\cite{Molnar:2003ff}. In this work, the NCQ scaling behavior of open-charm hadrons in p--Pb collisions is investigate for the first time. Figure~\ref{Fig: D_NCQ_v2} presents the $v_{2}/n_{\mathrm{q}}$ as the function of $kE_{\mathrm{T}}/n_{\mathrm{q}}$ for $D^{0}$, $D^{+}$, $D_s^{+}$ and $\Lambda_{c}^{+}$, and the comparison with the results of light-flavor hadron including $K_{s}^{0}$, $\pi^{\pm}$, $K^{\pm}$ is also shown. Compared to the Fig.~\ref{Fig: D_PID_v2}, the $v_{2}$ of all particle species is divided by the number of constituent quark $n_q$ ($n_q$ = 2 for meson, $n_q$ = 3 for baryon), and the $\it{p}_{\mathrm{T}}$ is replaced by the $n_{q}$-scaled transverse kinetic energy $kE_{\mathrm{T}}/n_{q}$ in consideration of the different mass of hadrons, where $kE_{\mathrm{T}} = m_{\mathrm{T}} - m_0 =\sqrt{p_{\mathrm{T}}^{2}+m_0^{2}-m_{0}}$. We found that all charm hadrons show a set of similar $v_{2}$ values after the NCQ scaling, confirming that the quark degree of freedom in flowing matter can also be probed for heavy quark in the transport model. On the other hand, the $v_{2}$ after the NCQ scaling for $K_s^{0}$ obtained with the parton scattering cross section $\sigma=0.5$ mb is compatible to results of $\pi^{\pm}$ and $K^{\pm}$, and all of them shows a larger value than charm hadrons. It suggests that the weaker collective behavior of charm quark compared to light quarks is mainly attributed to their different parton scatterings.

\section{Summary}
In summary, we studied the elliptic anisotropy of open charm hadrons in p--Pb collisions at $\sqrt{s_{\mathrm{NN}}}$ 8.16 TeV by means of introducing an additional charm quark–-antiquark pair production trigger in the AMPT model. The implementation of such trigger provides an efficient way to simultaneously describe the $p_{\mathrm{T}}$ spectrum and $v_{2}$ of $D^{0}$. Then we systematically investigated the dependence of $v_{2}$ on parton cross section in various multiplicity ranges, and demonstrate the importance of parton interactions for generating the collectivity of heavy quark in p--Pb collisions. In addition, we provided new predictions for the $v_{2}$ of other charm hadrons including $D^{+}$, $D_{s}^{+}$ and $\Lambda_{c}^{+}$ in p--Pb collisions. We argue that, the $v_{2}$ of open-charm hadron follows the NCQ scaling in high-multiplicity p--Pb collisions at LHC energies with a proper parton cross section value, indicating the existence of partonic degrees of freedom for heavy quarks in high-multiplicity small collision systems. Future studies about more types of heavy-flavor hadron, including charmonium and bottom hadron, can provide further understandings on the transport properties of heavy quark, then help to search for the possible formation of the hot medium in small systems.          

\section{Acknowledgement}
This work was supported by Natural Science Foundation of Hubei Provincial Education Department under Grant (Q20131603) and Key Laboratory of Quark and Lepton Physics (MOE) in Central China Normal University under Grant (QLPL2022P01, QLPL202106).

\bibliographystyle{apsrev4-2} 
\bibliography{reference}





\end{CJK}
\end{document}